# Modeling the Travel Time–Pollution Exposure Trade-off in Urban Route Choice

Cafer Avci[a], Zhaoyao Bao[b], Maximilian Dittgen[c], Danni Lu[a], Alireza Yazdiani[b], Mohammad Tayarani[b], H. Oliver Gao[a, d, *]

[a] Systems Engineering Program, Cornell University, Ithaca, NY, 14853, USA

[b] School of Civil and Environmental Engineering, Cornell University, Ithaca, NY, 14853, USA

[c] College of Computing and Information Science, Cornell University, Ithaca, NY, 14853, USA

[d] Center for Transportation, Environment, and Community Health (CTECH), Cornell University, 313 Hollister Hall, Ithaca, NY, 14853-3801, USA

## Abstract

Traffic-related air pollution is a major contributor to respiratory and cardiovascular disease, and travel time disproportionately drives daily exposure. Conventional routing algorithms minimize time or distance, ignoring street-level pollutant variability and leaving travelers exposed to avoidable risk. We develop a bi-objective routing framework that couples high-resolution AERMOD-based PM2.5 dispersion surface with a shortest-path algorithm using a deterministic, gradient-based weight-update scheme to enumerate Pareto-optimal paths along the time-exposure frontier. Applied to New York City's road network across 790,000 origin–destination pairs covering 1.48 million daily commuters, link-level concentrations differed by a factor of two or more between adjacent parallel streets. Healthy routes cut cumulative exposure by about 38% at a median time cost of about 28%, running in 30 milliseconds per query and recovering more frontier paths than uniform weight sampling (4.5 vs. 2.4). Results show exposure-aware routing is computationally practical at metropolitan scale, suitable for real-time navigation and infrastructure planning.

# 1. Introduction

Air pollution exposure in urban environments remains a critical public health issue, strongly associated with adverse outcomes such as respiratory and cardiovascular diseases and premature mortality (Beelen et al., 2014; Kim et al., 2015). Traffic-related air pollution (Harrison et al.) is particularly concerning due to its proximity to where people live, work, and travel, and its composition which includes fine particulate matter ($PM_{2.5}$), nitrogen oxides ($NO_x$), and brake and tire wear particles (Zhang and Batterman, 2013). While national and regional policies such as vehicle emission standards, fuel quality regulations, and the promotion of electric vehicles have contributed to emission reductions (Auffhammer and Kellogg, 2011; Cook et al., 2011; Jacobsen et al., 2023; Nopmongcol et al., 2017; Wallington et al., 2022), TRAP concentrations in urban areas often remain above safe levels, especially for vulnerable populations like children and the older people (Patton et al., 2025). Relying solely on long-term emission reduction scenarios without addressing immediate exposure risks may leave many urban residents vulnerable in the short term. Surprisingly, less attention has been paid to strategies that reduce individual exposure through behavioral or spatial interventions.

Travel time, although often brief, can contribute disproportionately to an individual's daily exposure (Tayarani and Rowangould, 2020). For example, although children spend only about 6% of their day commuting, the BREATHE project found that this brief window accounts for nearly 20% of their daily black carbon dose (Rivas et al., 2016). This underscores a critical gap in how exposure is distributed across time and space, which is often overlooked in conventional environmental impact assessment of transportation plans as well as epidemiology studies.

Traditional routing algorithms prioritize efficiency metrics such as travel time or distance, with little regard for environmental health impacts.

There is increasing awareness that routing strategies should go beyond simply optimizing speed and should also account for exposure to pollutants, especially for vulnerable populations who are more susceptible to air pollution. Recent efforts have begun to rethink how routing algorithms can reduce exposure to air pollution (Jiang et al., 2024). Gao et al., (2022), for instance, developed a pedestrian navigation tool that combines a neural network-based $PM_{2.5}$ estimator with a modified Dijkstra algorithm, resulting in up to a 20% reduction in exposure. Similarly, Zou et al., (2020) introduced the Healthier Route Planning (HRP) algorithm, which uses real-time pollution data to dynamically reroute travelers based on current air quality conditions. Commercial platforms like AirTrack developed by AirAware Labs have also been introduced, signaling both technical feasibility and growing public interest in healthier routing options.

A central challenge in exposure-aware routing is obtaining sufficiently granular air pollution data. Many routing algorithms rely on sparse monitoring networks that fail to capture the fine-scale spatial variability of urban air pollution. Vehicle-related pollutant levels are highly heterogeneous and can vary by orders of magnitude across short distances depending on traffic density, vehicle fleet composition, roadway geometry, travel speed, and meteorological factors such as wind and atmospheric mixing (Khuzestani et al., 2022). High-resolution exposure modeling requires the integration of traffic, meteorological, and urban morphology data. Dispersion models such as the U.S. EPA's AERMOD offer a physically grounded approach to estimating pollutant concentrations at fine spatial resolution by simulating atmospheric transport from emission sources to nearby receptors. When coupled with detailed traffic emission inventories, such models can produce street-level pollution surfaces that are essential for meaningful exposure-aware routing.

On the computational side, real-time routing at city scale requires the solution algorithm to be highly efficient. (Manja et al., 2024) compared heuristic-based A* algorithms with graph neural networks (GNNs) for health-optimal routing and found that while GNNs offered improved path quality, they struggled with scalability due to the computational overhead of repeated graph aggregation in dynamic environments. This highlights the need to balance the trade-off between model complexity and operational feasibility in real-time systems.

Reducing in-trip exposure matters because the relationship between $PM_{2.5}$ and adverse health outcomes is well documented in the epidemiology literature, typically expressed through concentration-response functions that relate incremental changes in pollutant concentration to changes in mortality or morbidity risk (Krewski et al., 2009; Pope et al., 2011). For instance, (Jerrett et al., 2008) found that reducing TRAP exposure by 20% among children was associated with a 30% reduction in asthma symptoms. Establishing how much exposure reduction is achievable through route choice is therefore a necessary first step. Translating those reductions into quantified health outcomes depends strongly on local ambient concentrations and baseline disease burden and is likely to be most consequential in high-pollution urban settings.

This study makes the following contributions:

1. We develop a dual-objective Pareto-frontier routing algorithm that balances travel time and TRAP exposure. The algorithm uses a deterministic weight-update scheme based on objective gradients to efficiently enumerate paths on the convex hull of the Pareto frontier, and we prove that each generated path is Pareto-optimal.

2. We integrate high-resolution $PM_{2.5}$ dispersion surfaces generated by the U.S. EPA's AERMOD model into a street-level routing framework and demonstrate its scalability in New York City, one of the largest and most complex urban networks in the world.
3. We quantify the exposure–time trade-off available to travelers at metropolitan scale, showing that healthy routes reduce cumulative $PM_{2.5}$ exposure by approximately 38% at a median travel-time cost of about 28%, and characterize how this trade-off varies across origin–destination pairs.

The remainder of this paper is organized as follows. Section 2 presents the problem formulation, including the air pollutant dispersion modeling framework and the bi-objective routing problem definition. Section 3 describes the solution algorithm, including the weighted-sum scalarization approach and the recursive Pareto-frontier generation procedure. Section 4 presents the case study design and numerical analysis, covering algorithm computational performance, air pollution modeling results, and routing results. Section 5 discusses the findings, and Section 6 concludes the paper.

# 2. Problem formulation

## 2.1. Air pollutant dispersion modeling

To estimate street-level $PM_{2.5}$ concentrations from vehicle traffic, we employ the U.S. EPA's AERMOD dispersion model. AERMOD is a steady-state Gaussian plume model that simulates the atmospheric transport and dispersion of pollutants emitted from near-surface sources. The model accounts for vertical variations in wind speed, turbulence, and temperature using boundary-layer similarity theory, and it distinguishes between stable and convective atmospheric conditions to determine plume rise, spread, and ground-level concentrations. These physical foundations make AERMOD well-suited for modeling the localized concentration gradients that characterize traffic-related air pollution in urban environments.

### 2.1.1 Input data

Three categories of input data are required. First, emission inputs were generated using the Cornell University Transportation and Environment Analysis Model (CU-TEAM), which estimates link-level $PM_{2.5}$ emission rates based on traffic volumes, vehicle speeds, and fleet composition derived from the regional travel demand model. Second, meteorological data were obtained from the NOAA Integrated Global Radiosonde Archive (IGRA) dataset (Durre et al., 2016), providing hourly surface observations (wind speed, wind direction, temperature, cloud cover) and upper-air soundings (mixing height, vertical temperature profiles) needed to characterize atmospheric stability and turbulence. Third, terrain elevation data from the U.S. Environmental Protection Agency (2024) were used to account for topographic effects on plume dispersion.

### 2.1.2 Modeling approach

In AERMOD, road segments carrying modeled traffic emissions are represented as line sources. Given the scale of the network and the number of receptor locations (approximately 1.96 million receptors at 100-meter resolution across the city), direct modeling of every source–receptor pair would be computationally infeasible. To address this, we adopted the raster-based subdomain decomposition method of Tayarani et al., (2018), which partitions the modeling domain into smaller subregions, each with a local receptor grid. AERMOD is run independently for each subdomain, and the resulting concentration fields are assembled into a city-wide pollution surface.

#### 2.1.3 Output

The model produces hourly $PM_{2.5}$concentration estimates at each receptor location. These point estimates are interpolated into fine-scale rasters representing the spatial distribution of traffic-related $PM_{2.5}$across the city. The $PM_{2.5}$concentration on each roadway link at each time of day is then estimated by performing a spatial join in ArcGIS Pro, overlaying the road network onto the concentration surface. The resulting link-level concentration values serve as the $AQI_a$ inputs to the routing problem defined in Section 2.2. The modeled concentrations were validated against measurements at air quality monitoring stations, showing strong correlations (Tayarani et al., 2020).

### 2.2. Bi-Objective Routing Problem

We consider a directed network graph $\boldsymbol{G} = (\boldsymbol{N}, \boldsymbol{A})$, where $\boldsymbol{N}$ is the set of nodes and $\boldsymbol{A}$ is the set of links (road segments). For each link $\boldsymbol{a} \in \boldsymbol{A}$, two attributes are defined: the travel time $t_{\mathrm{a}}(x_{\mathrm{a}})$, which depends on the traffic flow $x_{\mathrm{a}}$, and the air quality index $AQI_{\mathrm{a}}$, which represents the $PM_{2.5}$ concentration on link $a$ obtained from the dispersion model described in Section 2.1.

We define the exposure on link, $E_\alpha(x_\alpha)$, $\alpha$ as the product of concentration and travel time:

$$E_\alpha(x_\alpha) = AQI_\alpha \cdot t_\alpha (x_\alpha) \qquad \text{Eq. (1)}$$

For a path P connecting origin r to destination s, the total travel time and total exposure are:

$$T(P) = \sum_{a \in \mathrm{P}} t_\alpha\ (x_\alpha) \qquad \text{Eq. (2)}$$

$$E(P) = \sum_{a \in \mathrm{P}} E_\alpha\ (x_\alpha) \qquad \text{Eq. (3)}$$

The problem is subject to the following constraints:

1. Flow conservation: $\sum_k f_{rs}^k = q_{rs}, \forall r, s$ (Eq. 4) where $f_{kr}{}^{s}$ is the flow on path k between origin r and destination s, and $q_{rs}$ is the travel demand between r and s.
2. Link flow–path flow relationship: $x_\alpha = \sum_{r,s} \sum_k \delta_a^{k,rs} f_{rs}^k, \forall \alpha$ (Eq. 5) where $\delta_{\mathrm{a}}{}^{\mathrm{k} \cdot \mathrm{rs}}$ is a binary indicator equal to 1 if link $a$ belongs to path k for the pair $(r, s)$.
3. Non-negativity: $f_{rs}^k \geq 0,\ \forall k, r, s\ (Eq. 6)$

The multi-objective shortest path algorithm ensures that no traveler can reduce their exposure to TRAP by unilaterally changing their route. This is expressed as:

$$E_{rs}^k = \min\{E_{rs}^k\}, \forall k, r, s \qquad \text{Eq. (7)}$$

Where:

$$E_{rs}^k = \sum_\alpha E_\alpha(x_\alpha) \delta_\alpha^{k,rs} \qquad \text{Eq. (8)}$$

And $E_{rs}^k$ is the total exposure on path $k$ from origin $r$ to destination $s$.

We assume that link travel times and pollution concentrations are exogenous for a given time period (i.e., we do not model congestion feedback from rerouting). Under this assumption, the bi-objective problem reduces to finding, for each OD pair independently, the set of Pareto-optimal paths with respect to time and exposure.

## 3. Solution algorithm

The bi-objective routing problem defined in Section 2.2 cannot be solved directly by standard single-objective shortest path algorithms, which assume a single cost function (typically travel time) on each link. Furthermore, precomputation-based speedup techniques widely used in practice (Delling et al., 2009) assume static link costs, whereas in our setting the relative weighting of time and exposure may vary across queries. We therefore adopt a weighted-sum scalarization approach that reduces the bi-objective problem to a sequence of single-objective shortest path problems, each solvable by Dijkstra's algorithm.

### 3.1.Weighted-Sum Scalarization

For a parameter α ∈ [0, 1], we define a composite link cost:

$$C(x_\alpha, \alpha) = t(x_a) \cdot (1 - \alpha + \alpha \cdot AQIx_\alpha) \qquad \text{Eq. (9)}$$

When $\alpha = 0$, the cost reduces to travel time alone; when $\alpha = 1$, the cost equals the exposure $E_\mathrm{a}(x_\mathrm{a})$. For intermediate values, the cost represents a blend of the two objectives. We show that any path minimizing $C(x_\mathrm{a}, \alpha)$ is Pareto-optimal with respect to time and exposure (Proof 1, Appendix). This property ensures that by varying α, we trace paths along the convex hull of the Pareto frontier.

This is not to say that every α will produce a unique path, or that we will be able to find every Pareto-optimal path, paths on non-convex portions of the frontier remain hidden. Rather, we can associate each path on the convex hull of the Pareto frontier with a range of α values that generate that path.

### 3.2. Recursive Pareto-Frontier Generation

To enumerate all discoverable Pareto-optimal paths for a given OD pair, we begin by solving for the two extreme paths: the time-optimal path (α = 0) and the exposure-optimal path (α = 1). If these paths differ, we recursively search for intermediate paths by computing a new α value based on the gradient between two known adjacent paths on the frontier:

$$\alpha' = 1 - \frac{|path_1.exposure - path_0.exposure|}{|path_1.travel_{time} - path_0.travel_{time}|} \qquad \text{Eq.}(10)$$

This deterministic weight-update strategy targets regions of the frontier where new paths are most likely to exist, rather than uniformly sampling α values. The recursion terminates when no new unique path is found between two adjacent known paths. Algorithms 1 and 2 present the initialization and recursive collection procedures, respectively.

**Algorithm 1** Find Pareto-Optimal Paths

$Procedure\ FindParetoOptimalPaths(G, start_node, end_node)$

  $graph_0 \leftarrow UpdateEdgeCosts(G, a = 0)^{+}$

  $path_0 \leftarrow ShortestPath(graph_0, start_node, end_node)$

  $unique_{paths}.add(path_0)$

  $graph_1 \leftarrow UpdateEdgeCosts(G, a = 1)$

  $path_1 \leftarrow ShortestPath(graph_1, start_node, end_node)$

  $\boldsymbol{if}\ path_0 \neq path_1 \boldsymbol{then}$

    $unique_paths.add(path_1)$

    $FindIntermediatePaths(G, unique_paths, path_0, path_1)$

  $\boldsymbol{end\ if}$

```
        return unique_paths
    end procedure
```

**Algorithm 2** Recursively Find Intermediate Paths

```
Procedure FindIntermediatePaths(G, unique_paths, path_0, path_1)
    gradient ← |path_1.exposure − path_0.exposure| / |path_1.travel_time − path_0.travel_time|
    a ← 1 − gradient
    graph' ← UpdateEdgeCosts(G, a)
    path' ← ShortestPath(graph', start_node, end_node)
    if path' ∉ unique_paths then
        unique_paths.add(path_1)
        FindIntermediatePaths(G, unique_paths, path_0, path')
        FindIntermediatePaths(G, unique_paths, path', path_1)
    end if
end procedure
```

# 4. Numerical analysis

## 4.1. Case Study and Experimental Design

New York City was selected as the experimental domain due to its comprehensive traffic network and trip generation data, as well as its significant heterogeneity in both route options and exposure levels between origin-destination pairs. This variability creates an ideal environment for examining large-scale aggregated exposure and travel time trade-offs across the Pareto frontier.

We accessed hourly traffic-related $PM_{2.5}$ concentration data for New York City's major arterials and local roadways from the AERMOD model described in Section 2.1. The concentration data were interpolated onto the traffic network via spatial join in ArcGIS Pro, aligning $PM_{2.5}$values with each road network link. This network serves as the input graph for the Pareto-optimal path-finding algorithm defined in Section 3.

Travel demand data for each OD pair were derived from the trip distribution step of the New York Best Practice Model (NYBPM) activity-based model developed by the New York Metropolitan Transportation Council (New York Metropolitan Transportation Council (NYMTC)). Using daily road traffic trip tables, we created a corpus of 10,000 origin-destination pairs randomly sampled from the distribution of observed commuter trips. Each trip is represented as a tuple containing an origin and destination TAZ, with a randomly assigned coordinate within each TAZ serving as the trip endpoint.

We use the Pareto-frontier algorithm to generate routes for each of the 10,000 OD pairs under two experimental setups designed to test both the weight-update strategy and the sensitivity to time of day:

1. 9 AM $PM_{2.5}$ concentration levels with the deterministic gradient-based weight update.
2. 5 PM $PM_{2.5}$ concentration levels with a uniform linspace weight update (25 values of α between 0 and 1).

For the full-scale analysis, we sample one random point in each of 3,585 New York City TAZs and construct all ordered OD combinations between distinct TAZs, yielding $3{,}585^2 - 3{,}585$ representative inter-TAZ OD pairs. For each pair, we generate Pareto-frontier paths and compute the difference in travel time and cumulative $PM_{2.5}$ exposure between the lowest-time and lowest-exposure routes, weighted by the number of commuters travelling that OD pair per day.

**Table 1.** Overview of the Dataset Used for the Experiment

| **Number of commuters accounted for** | **1,478,975** |
|---|---|
| Total number of O-D pairs processed | 790,103 |
| Routes generated per O-D pair (Mean, Med, Std) | 5.47, 5.00, 3.22 |

### 4.2. Algorithm Computational Performance

We first evaluate the computational performance of the proposed algorithm on a network graph of Miami, Florida, consisting of 7,082 edges and 4,103 nodes. After generating routes between 1,000 random origin-destination pairs at least 1 km apart, the algorithm produced an average of 3.44 Pareto-optimal routes per pair (median: 3). Each route generation took an average of 30 milliseconds on a single core of an Apple M2 Pro processor (median: 25 ms), confirming that the approach is suitable for real-time applications.

The elasticity of travel time with respect to $PM_{2.5}$exposure was $-6.80$ on average for the 1,000 pairs (median: $-2.69$), indicating that small increases in travel time can yield large reductions in exposure. The average difference in travel time between the fastest and healthiest routes was 21.5% (median: 15.24%), while the average difference in exposure was 56.1% (median: 43.6%).

### 4.3. Air Pollution Modeling Results

Figure 1 presents the average daily traffic-related $PM_{2.5}$ concentration surface produced by AERMOD for the New York City study area. The map reveals substantial spatial heterogeneity, with elevated concentrations along major highways and arterials and substantially lower concentrations on residential side streets.

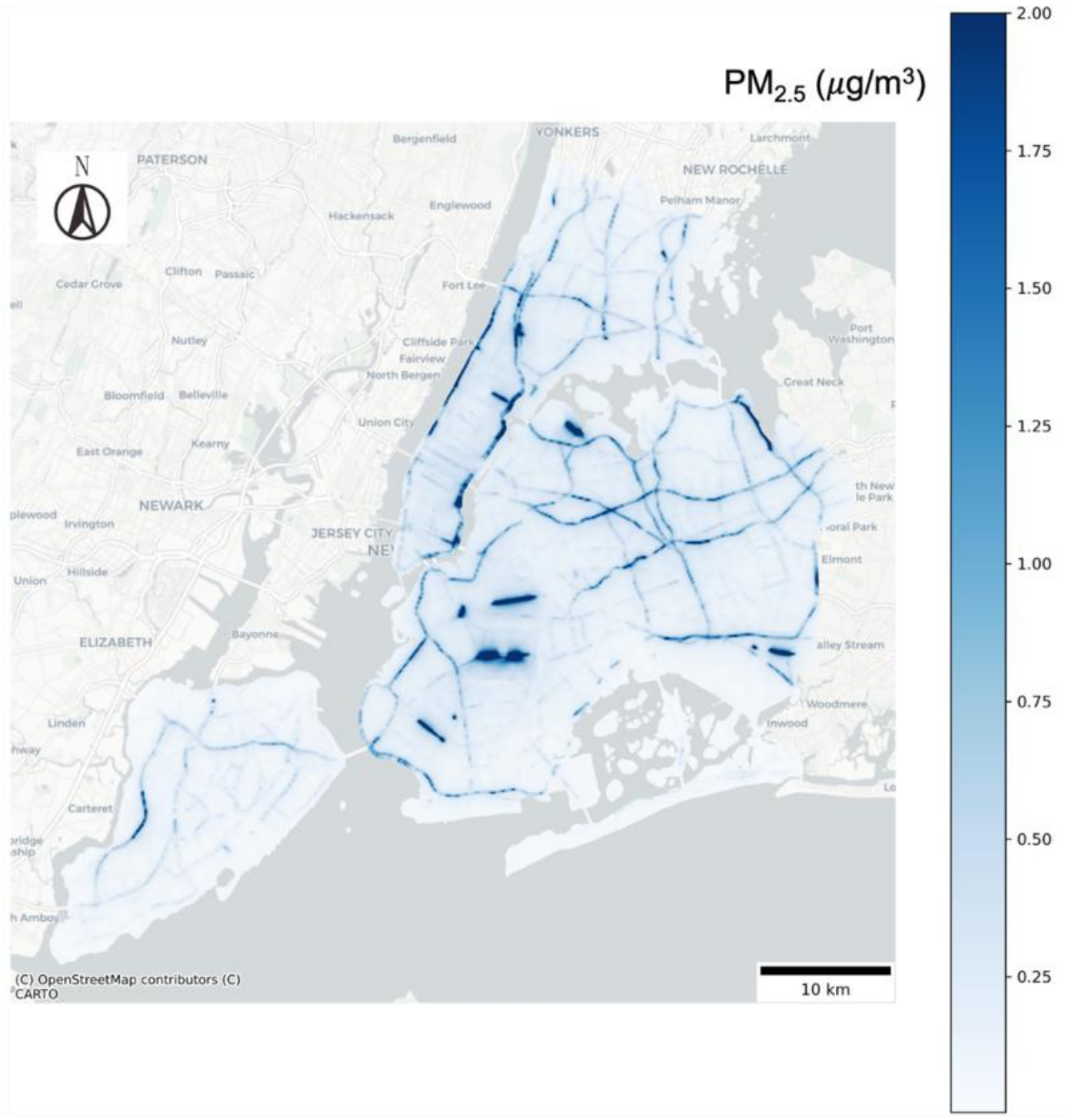

**Figure 1.** Average Daily Traffic-Related $PM_{2.5}$Concentration Output from U.S. EPA AERMOD.

To illustrate the fine-scale variability that the routing algorithm exploits, Figure 2 presents a zoomed-in view of link-level $PM_{2.5}$concentrations for Manhattan. The street-level map shows that even within a small area, adjacent parallel streets can differ in $PM_{2.5}$concentration by a factor of two or more, creating meaningful opportunities for exposure reduction through route choice.

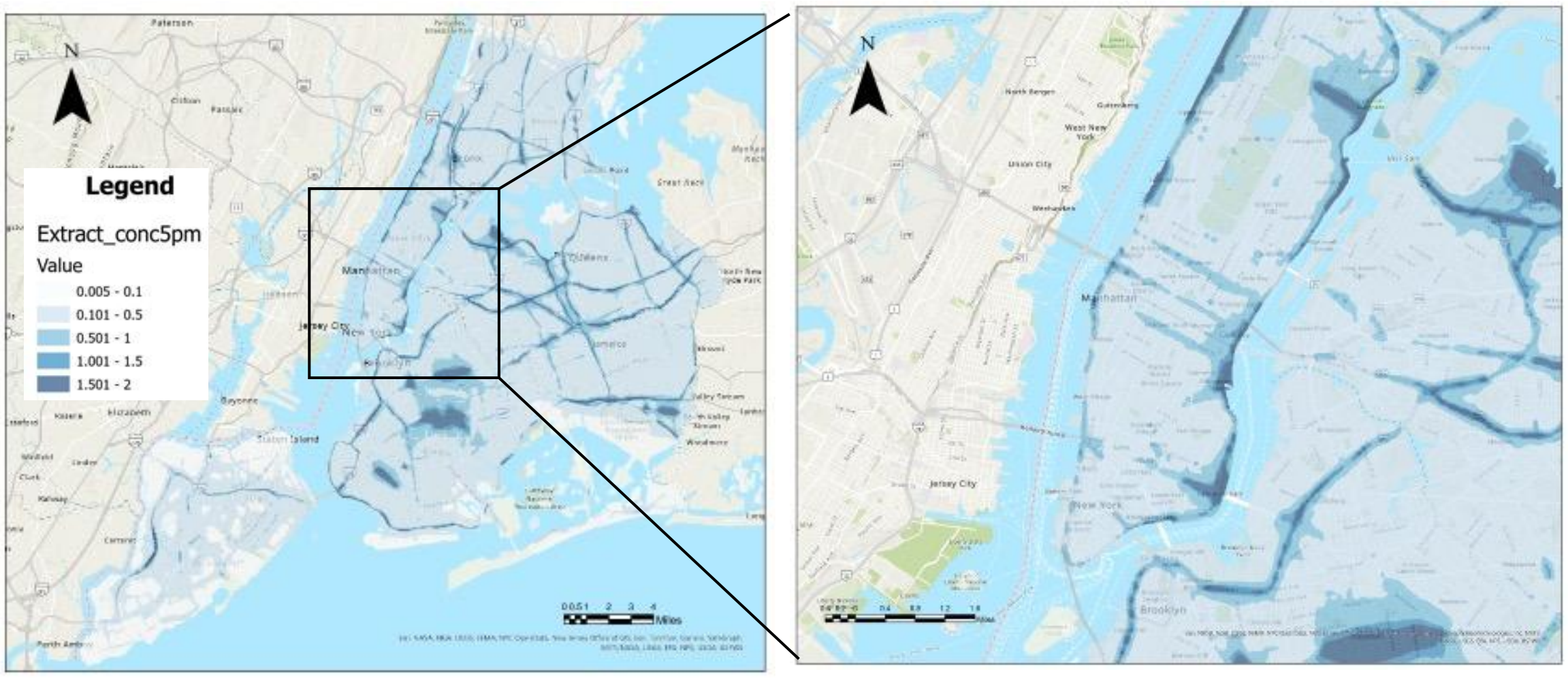

**Figure 2.** Street-Level $PM_{2.5}$Concentrations for Manhattan.

Table 2. summarizes the distribution of link-level $PM_{2.5}$concentrations across the network for both the 9 AM and 5 PM time periods.

**Table 2.** Distribution of Link-Level Traffic-Related $PM_{2.5}$Concentrations (µg/m³)

| **Statistic** | **9 AM** | **5 PM** |
|---|---|---|
| Mean | 0.2258 | 0.3689 |
| Median | 0.1098 | 0.2364 |
| Standard Deviation | 0.7387 | 0.8333 |
| 25th Percentile | 0.0562 | 0.1311 |
| 75th Percentile | 0.2049 | 0.3858 |
| Minimum | 0.0018 | 0.0037 |
| Maximum | 2.4652 | 2.5810 |

The 9 AM period shows lower average concentrations than the 5 PM period, consistent with diurnal traffic patterns and previous findings of lower overall $PM_{2.5}$ concentration during morning hours (Desai et al., 2022). The large spread between the 25th and 75th percentiles confirm that significant exposure variation exists across the network, providing the basis for exposure-aware route differentiation.

### 4.4. Routing Results

The dual-objective routing experiments show that healthy, low-exposure routes can substantially reduce in-vehicle $PM_{2.5}$ exposure at the cost of moderate increases in travel time.

Table 3 presents results from the 10,000-pair sample. The deterministic gradient-based update generated more Pareto-optimal paths per OD pair than the uniform linspace weighting (mean 4.51 vs. 2.40 paths; medians 3 vs. 1), confirming that exploiting objective gradients explores the frontier more effectively than uniform sampling. Deterministic runs also produced longer average travel times (1,350 s vs. 979 s) and higher variability in both time and exposure, consistent with sampling a wider range of extreme and intermediate routes.

At 9 AM, minimum-time (shortest) routes had relatively low average $PM_{2.5}$concentrations (mean 0.098 μg/m³, median 0.075 μg/m³), while at 5 PM minimum-exposure (healthiest) routes were cleaner (mean 0.089, median 0.083), reflecting diurnal differences in traffic-related pollution consistent with previous findings (Desai et al., 2022). In both time periods, shifting from the shortest route to the healthiest route reduced average route-level $PM_{2.5}$by about 28-34%, with corresponding travel time increases of about 42-47%.

Applying the deterministic approach to the full TAZ dataset ($3{,}585^2 - 3{,}585$ OD pairs), Table 4 shows that the median healthy route increased travel time by about 28–29% relative to the shortest path while reducing cumulative $PM_{2.5}$exposure by roughly 38-42%, implying an elasticity of approximately 3-4. These averages mask substantial heterogeneity: large standard deviations in both time and exposure differences indicate that some OD pairs offer very favorable exposure reductions for little extra time while others require larger time sacrifices.

**Table 3.** Characteristics of routes from the 10,000-pair sample under the deterministic (9 AM) and linspace (5 PM) weight-update strategies.

| **Metric** | | **9am Deterministic** | **5pm Linspace** |
|---|---|---|---|
| Paths per route | Mean | 4.51 | 2.40 |
| | Median | 3.00 | 1.00 |
| | SD | 4.36 | 2.45 |
| Route time (seconds) | Mean | 1350.23 | 979 |
| | Median | 915.18 | 641 |
| | SD | 1159.84 | 1071 |
| Average $PM_{2.5}$ Concentration $^{\mu g}/_{m^3}$ Along Min-Travel Time Route | Mean | 0.0978 | 0.2304 |
| | Median | 0.0752 | 0.1545 |
| | SD | 0.0881 | 0.3757 |
| Average $PM_{2.5}$ Concentration $^{\mu g}/_{m^3}$ Along Min-Exposure Route | Mean | 0.1591 | 0.0891 |
| | Median | 0.1068 | 0.0834 |
| | SD | 0.0881 | 0.3757 |
| Difference in Travel Time, Min-Travel Time vs Min Exposure Route | Mean | 47.16% | 41.82% |
| | Median | 28.65% | 28.56% |
| | SD | 48.68% | 44.89% |
| Difference in the exposure; Min Travel Time vs Min Exposure Route | Mean | 34.03% | 28.28% |
| | Median | 28.08% | 25.47% |
| | SD | 21.59% | 20.21% |
| Elasticity of Exposure Route to Travel Time | Mean | 2.01 | 2.09 |
| | Median | 0.89 | 0.74 |
| | SD | 6.51 | 13.08 |

**Table 4.** Characteristics of the shortest path routes and exposures for the healthy routes

| **Route Overview** | **Mean (SD)** | **Median** |
|---|---|---|
| Travel time (seconds) | 691.5 (381.7) | 628.9 |
| Exposure ($PM_{2.5}$ ($^{\mu g}/_{m^3}$)*seconds) | 132.02 (119.95) | 96.45 |
| % Difference in travel time (healthy vs shortest) | 33.7% (27.21%) | 27.59% |
| % Difference in pollution (healthy vs shortest) | 38.45% (21.86%) | 41.72% |
| Elasticity of pollution w.r.t. time | 3.79 (164.43) | 1.24 |

To illustrate the structure of the Pareto frontier for a single OD pair, Figure 3 shows the set of Pareto-optimal routes generated for a representative pair on the New York City road network. The paths are approximately evenly distributed along the frontier, demonstrating that the deterministic bisect-and-recurse weight-update strategy effectively explores the trade-off space between travel time and exposure.

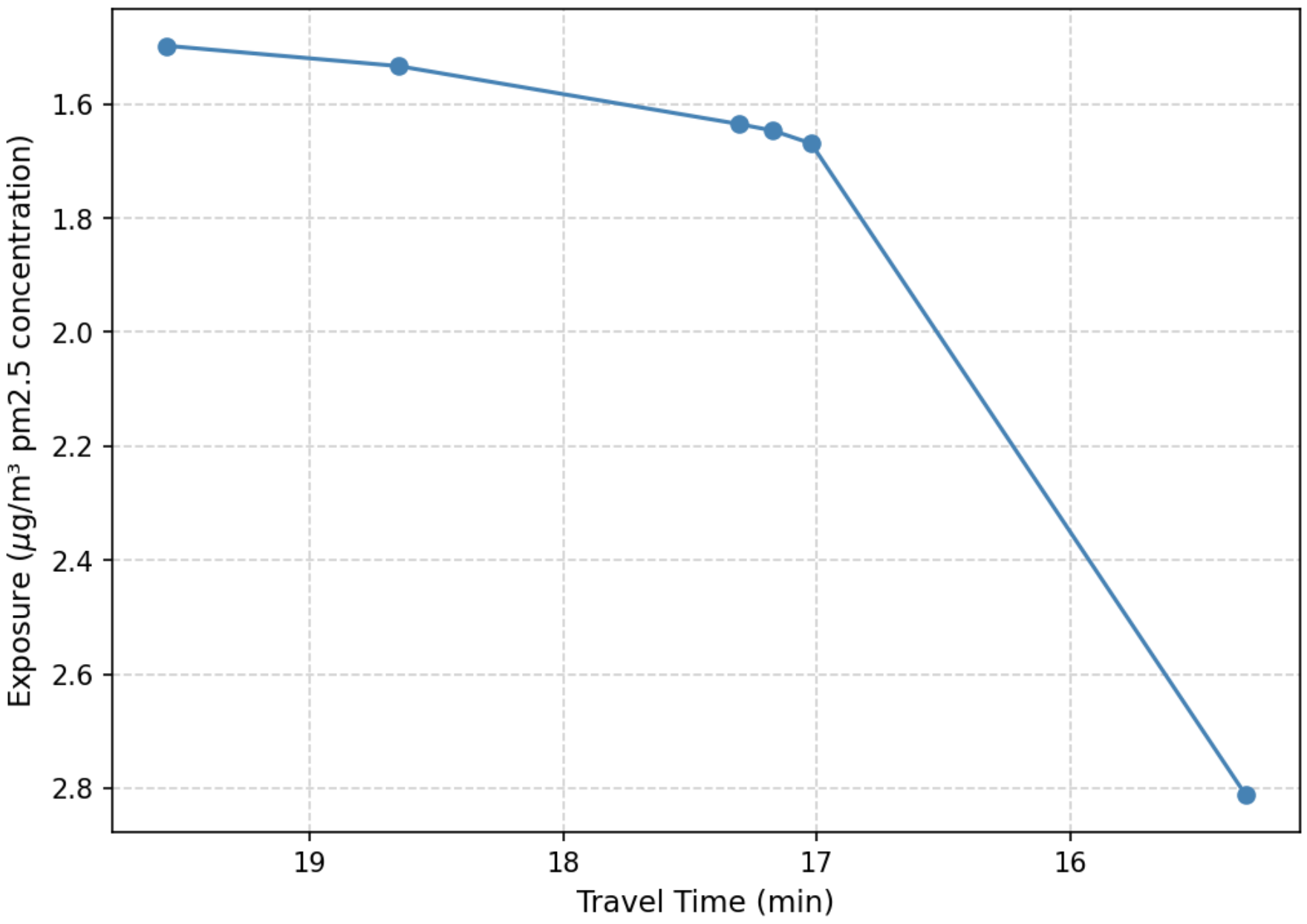

**Figure 3.** Sample Pareto Frontier for a 6-path set of routes between an origin/destination pair in New York, NY

Figure 4 depicts trip counts across road segments for both healthy and shortest-route scenarios. As shown in panel (c), healthy routes tend to favor local streets where pollution concentrations are lower, while shortest routes primarily use highways to minimize travel time by leveraging higher speeds.

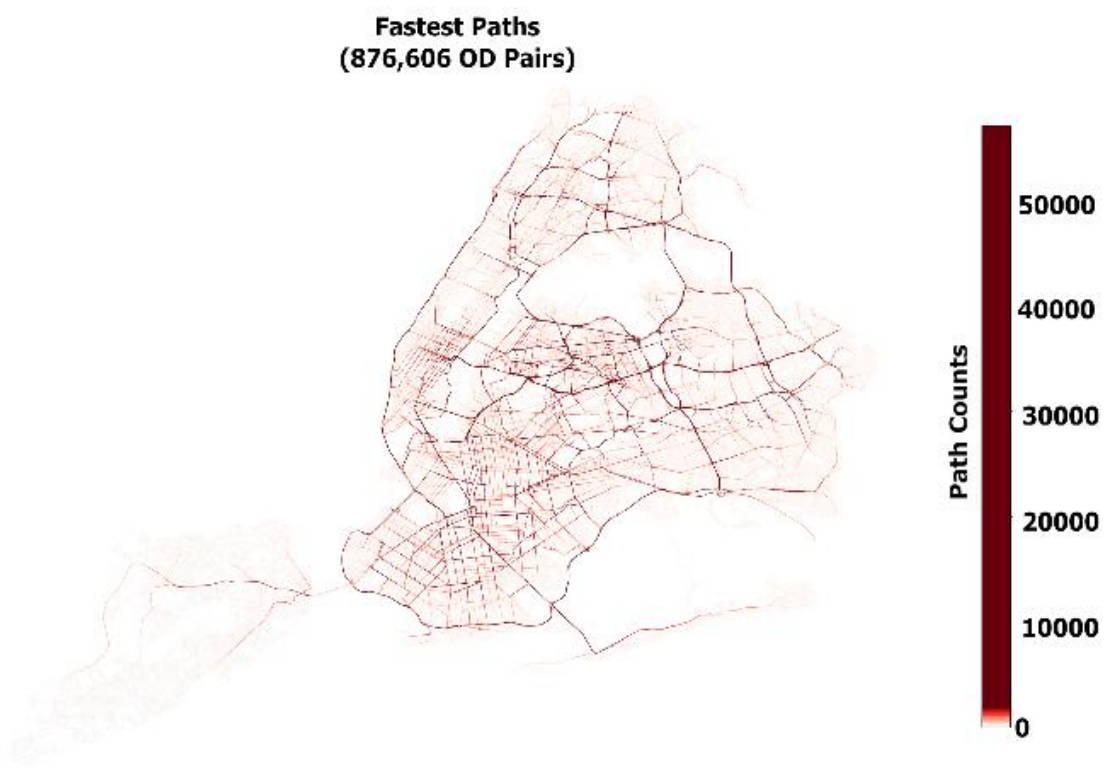

(a) Routes for short routes

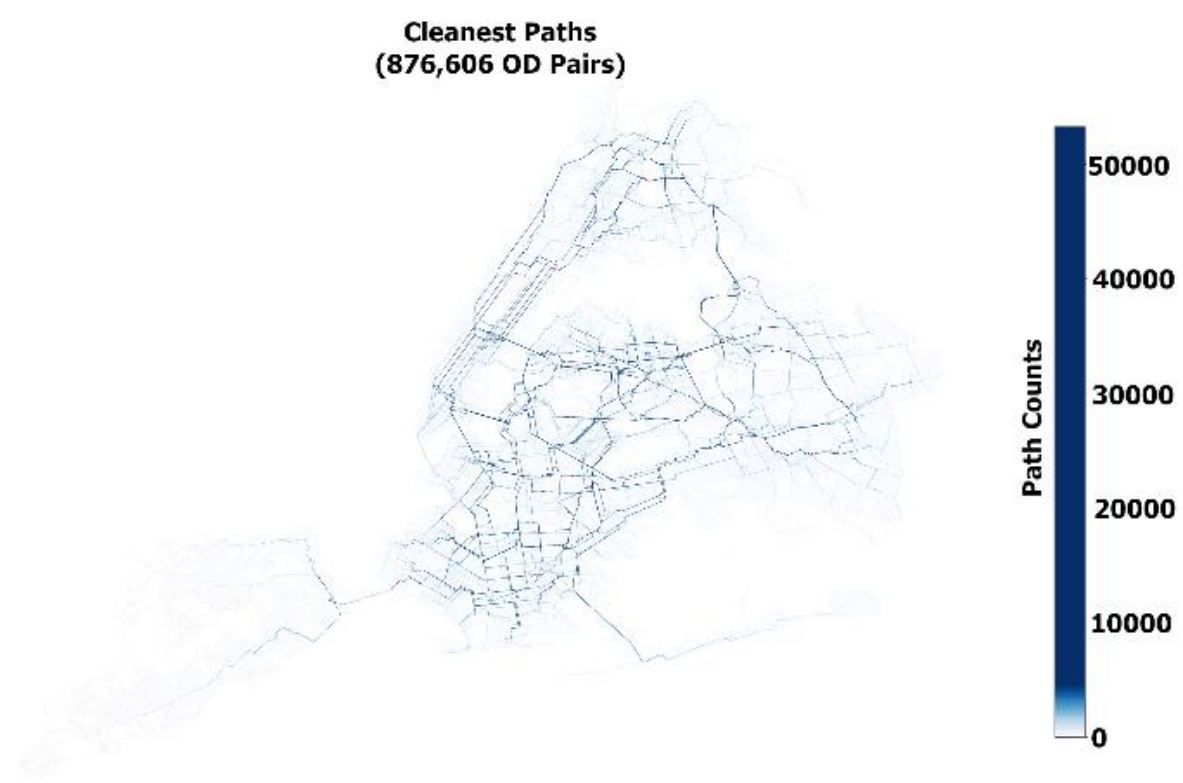

(b) Routes for healthy routes

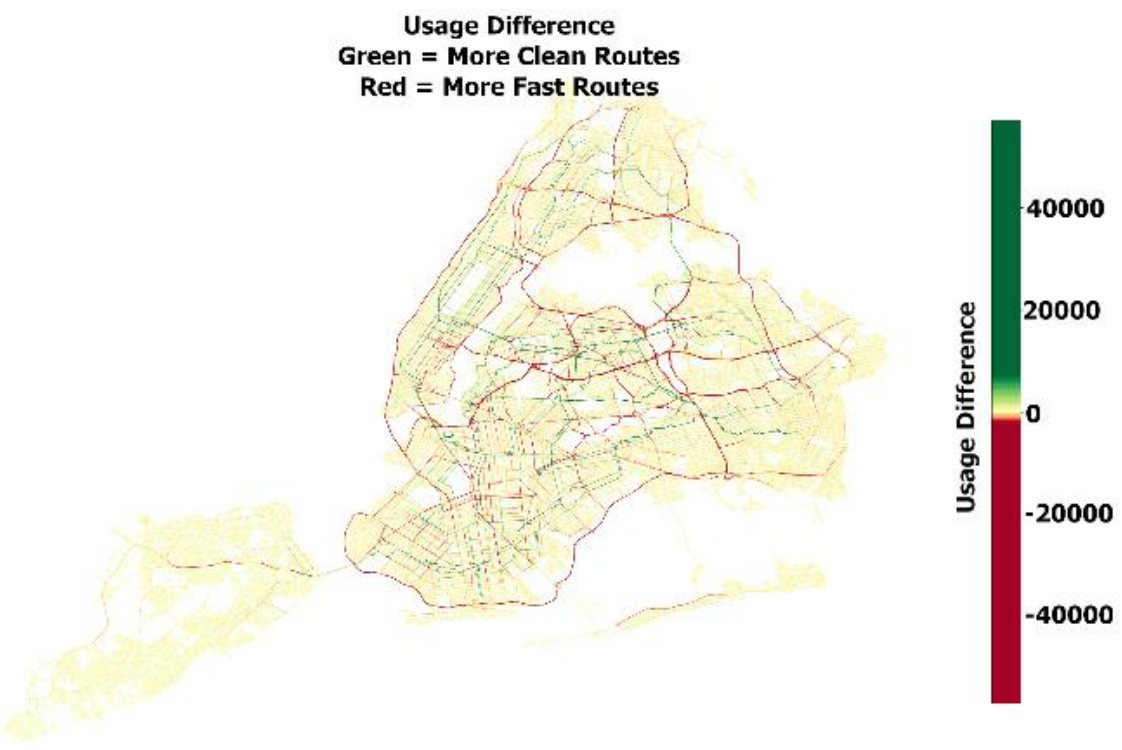

(c) Difference in trip counts per route under short and healthy routes

**Figure 4.** The trip counts for each road segment in the network under short and healthy routes and the differences

## 5. Discussion

Despite decades of regulatory efforts to reduce transportation-related emissions and increasing public awareness of the health impacts of traffic-related air pollution, people around the world continue to be exposed to harmful levels of pollutants and suffer adverse health outcomes (Brugge et al., 2007). One of the most overlooked yet significant sources of exposure is the time individuals spend traveling, when they are directly in contact with the highest concentrations of vehicle emissions (Setton et al., 2011; Tayarani and Rowangould, 2020).

A promising strategy to mitigate exposure is through exposure-aware routing. Instead of prioritizing the shortest or fastest route, travelers can be guided along paths with lower pollution concentrations. Vehicle-related pollutant levels are highly heterogeneous and can vary by orders of magnitude across short distances depending on traffic density, vehicle fleet composition, roadway geometry, travel speed, and meteorological factors such as wind and atmospheric mixing (Khuzestani et al., 2022). Traditional routing algorithms, however, almost exclusively optimize travel time, thereby neglecting potential health costs associated with pollution exposure.

Although travel time remains the most important factor influencing route choice (Ben-Akiva and Lerman, 1985; Chorus et al., 2007), studies suggest that secondary considerations such as safety, comfort, and health can also significantly shape decisions, especially when travelers are presented with clear information (Avineri and Prashker, 2005). Prior work has shown that individuals are more willing to adopt healthier or environmentally preferable routes when provided with quantitative feedback on benefits, particularly during non-peak hours when time constraints are less pressing (Siriaporn et al., 2025). These behavioral findings suggest that exposure-aware routing could realistically influence travel decisions if integrated into navigation platforms.

In this study, we developed and implemented a routing algorithm that minimizes TRAP exposure and applied it to New York City, one of the largest and most congested urban environments in the world. NYC provides a particularly rigorous test case due to its congested streets, dense street network, and significant spatial variation in pollution. Our results demonstrate that the algorithm is scalable and effective in large, real-world urban networks. Specifically, we found that the median healthy route reduced cumulative exposure by approximately 42% relative to the shortest path, at a median travel-time cost of about 28%. Whether this trade-off is worthwhile depends on how travelers value time against avoided exposure; future work could formalize it by monetizing both the avoided health damages (e.g., reduced hospitalizations, lower long-term cardiovascular and respiratory risks) and the opportunity costs of additional time spent traveling (Wang et al., 2024). These values are consistent with the full-scale analysis reported in Table 4, which shows substantial heterogeneity across OD pairs and indicates that some trips offer large exposure reductions for relatively small increases in travel time.

The practical applications of our routing algorithm are multifaceted. First, it can be integrated into standalone navigation tools that offer users route options optimized for reduced pollution exposure. Second, it can be embedded within mainstream platforms such as Google Maps, Apple Maps, or Waze, enabling users to compare traditional shortest routes with healthier alternatives. Beyond individual-level navigation, the algorithm has broader implications for urban planning and public health policy. Exposure-aware routing can inform the strategic design of active transportation infrastructure such as bike lanes and pedestrian corridors by prioritizing alignments that minimize exposure to traffic-related air pollutants. Given the high spatial and temporal variability of vehicular emissions, which differ significantly from the more uniform patterns of background pollution, this approach supports more nuanced transportation and land-use planning (Tayarani et al., 2016). Moreover, exposure-sensitive site selection for schools, senior housing, and healthcare facilities where vulnerable populations spend substantial time can yield significant public health benefits.

The street-level concentration results further demonstrate why this type of routing can be effective. Even adjacent parallel streets in Manhattan can differ in traffic-related $PM_{2.5}$ concentration by a factor of two or more, creating meaningful opportunities to reduce exposure through relatively small changes in route choice. The routing results also show that healthy routes tend to favor local streets with lower pollution concentrations, whereas shortest routes rely more heavily on highways to minimize travel time.

Several limitations frame these results. We treat link-level travel times and pollutant concentrations as exogenous; large-scale adoption of exposure-aware routing would redistribute traffic and alter both, an effect that would require equilibrium modeling to capture. We also model $PM_{2.5}$ alone, whereas travelers are simultaneously exposed to $NO_2$, ultrafine particles, and black carbon, which exhibit different spatial distributions and may not co-vary with $PM_{2.5}$ at the street level.

Most importantly, this study quantifies exposure reduction rather than downstream health outcomes. That choice reflects the study setting rather than a limitation of the framework itself. New York City has comparatively low ambient $PM_{2.5}$ concentrations by global standards, traffic contributes a modest share of total pollution-attributable mortality, and commuting accounts for only a fraction of daily exposure, so the absolute mortality reduction attributable to commute-time route choice alone is inherently small here. Reliable health quantification would also require concentration–response functions appropriate to short-duration exposure, whereas the available cohort literature is derived from long-term average exposure. Applying this framework in cities with substantially higher ambient concentrations, such as those in South and East Asia, would likely produce health benefits large enough to quantify meaningfully and is a priority for future work. These limitations and future directions are consistent with the scope of the current study, which focuses on quantifying achievable exposure reductions through route choice rather than estimating downstream health outcomes.

## 6. Conclusions

Overall, our results highlight that reframing routing as a multidimensional optimization problem can yield substantial reductions in traveler exposure to traffic-related air pollution. Applied at metropolitan scale in New York City, the proposed framework shows that healthy routes can reduce cumulative exposure by approximately 42% at a median travel-time cost of about 28%. The deterministic gradient-based weight-update strategy also generated more Pareto-optimal alternatives than uniform weight sampling, demonstrating that the proposed approach can efficiently characterize the trade-off between travel time and exposure.

The successful implementation in NYC demonstrates that such algorithms are both computationally feasible and deployable at metropolitan scale. The algorithm required approximately 30 milliseconds per route generation in the computational performance experiment, supporting its potential use in real-time applications. Beyond individual navigation, exposure-aware routing can support transportation planning and public health decision-making by identifying lower-exposure alternatives and informing the design of transportation infrastructure.

Future research should extend this work by incorporating dynamic pollution models, integrating real-time sensor data, exploring user adoption behavior, accounting for multiple pollutants, evaluating network-wide traffic responses to large-scale adoption, and examining the equity implications of exposure-aware routing. These extensions can further establish exposure-aware routing as a practical approach for reducing traffic-related air pollution exposure and supporting healthier urban mobility.

**Declaration of competing interest**

The authors declare that they have no known competing financial interests or personal relationships that could have appeared to influence the work reported in this paper.

**Data availability**

The travel-demand data used in this study was obtained from the regional metropolitan planning organization under a data-use agreement and cannot be shared publicly. Aggregated origin-destination and link-level exposure results, and the routing code, are available from the corresponding author upon reasonable request. AERMOD dispersion model configuration files are available from the authors upon request, subject to any third-party data restrictions.

# Appendix

## 1. Proof 1:

Proof. We prove that any $\alpha \in [0,1]$ the path $P_\alpha$ minimizing the weighted cost function $C(x_\alpha, \alpha) = t(x_\alpha).(1 - \alpha + \alpha.AQIx_\alpha)$ is Pareto-optimal.

Let $P_\alpha$ be the path returned by the shortest path algorithm for some $\alpha \in [0,1]$. Suppose, by contradiction, that $P_\alpha$ is not Pareto-optimal. Then there exists some path $P'$ such that:

$$t(P') \leq t(P_\alpha) and\ E(P') < E(P_\alpha)$$

or

$$t(P') < t(P_\alpha) and\ E(P') \leq E(P_\alpha)$$

For the weighted cost of $P_\alpha$, we have:

$$C(P_\alpha, \alpha) = (1 - \alpha)t(P_\alpha) + \alpha.E(P_\alpha)$$

And for $P'$:

$$C(P', \alpha) = (1 - \alpha)t(P') + \alpha.E(P')$$

Given either dominance condition, we can show:

$$\begin{aligned} C(P', \alpha) &= (1 - \alpha)t(P') + \alpha.E(P') \\ &< (1 - \alpha)t(P_\alpha) + \alpha.E(P_\alpha) \\ &= C(P_\alpha, \alpha) \end{aligned}$$

This contradicts our assumption that $P_\alpha$ minimizes the weighted cost function. Therefore, any path found using the weighted cost function must be Pareto-optimal.